\documentclass[12pt]{article}
\usepackage{latexsym}
\usepackage{graphicx}
\newcommand{\beq}{\begin{equation}}
\newcommand{\eeq}[1]{\label{#1}\end{equation}}
\newcommand{\bea}{\begin{eqnarray}}
\newcommand{\eea}[1]{\label{#1}\end{eqnarray}}

\begin{document}
\begin{flushright}
\hfill{IMPERIAL-TP-2014-MJD-05}\\
%\hfill{hep-th/0208093 }
\end{flushright}
\bigskip
\bigskip
%\fpage{1}
\centerline{\bf How fundamental are fundamental constants?\footnote{Review commissioned by Contemporary Physics. DOI:10.1080/00107514.2014.980093}}

 \vspace*{0.37truein} \centerline{\footnotesize M.
J.  Duff\footnote{m.duff@imperial.ac.uk }} \vspace*{0.015truein}
\centerline{\footnotesize\it Blackett Laboratory, Imperial College London, SW7 2AZ, UK}

\bigskip

\vspace{20pt}

\abstract{I argue that the laws of physics should be independent of one's choice of units or measuring apparatus. This is the case if they are framed in terms of dimensionless numbers such as the fine structure constant, $\alpha$.  For example, the Standard Model of particle physics has 19 such dimensionless parameters whose values all observers can agree on, irrespective of what clock, rulers, scales... they use to measure them. Dimensional constants, on the other hand, such as $\hbar$, $c$, $G$, $e$, $k$\ldots, are merely human constructs whose number and values differ from one choice of units to the next. In this sense only dimensionless constants are ``fundamental".  Similarly, the possible time variation of dimensionless fundamental ``constants'' of nature is operationally well-defined and a legitimate subject of physical enquiry.  By contrast, the time variation of dimensional constants such as $c$ or $G$ on which a good many (in my opinion, confusing) papers have been written, is a unit-dependent phenomenon on which different observers might disagree depending on their apparatus. All these confusions disappear if one asks only unit-independent questions.
 We provide a selection of opposing opinions in the literature and respond accordingly.

\newpage
%\tableofcontents
%\begin{document}
\newpage
\section{Dimensionless versus dimensional constants}
\indent

In physics it is important to distinguish between statements about the universe we live in and 
statements about the human conventions we adopt to describe it.  For example, ``The fine structure constant, $\alpha$, is approximately 1/137''  is a universal statement. As Richard Feynman \cite{Feynman1} put it ``You might say the {\it hand of God} wrote that number, and we don't know how He pushed his pencil.''}, whereas the statement ``The speed of light in vacuum, $c$, is $3 \times 10^{8}$ meters per second'' merely tells us how to convert one human construct, the meter, into another, the second.  Accordingly it is matter of convention whether $c$ is something we measure or something we define to be fixed. Indeed, the  \emph{Conf\'{e}rence G\'{e}n\'{e}rale des Poids et Mesures} (CGPM) effectively varied  the speed of light (VSL) when it adopted the latter convention in 1982.  The Universe carried on regardless.

We shall argue in section 2, therefore, that any ``law of nature'' worth its salt should be universal in this sense, and one upon which everyone will agree irrespective of their choice of units or choice of measuring apparatus. This is easily achieved by framing our laws in terms of dimensionless constants such as $\alpha$ rather than dimensional ones such as the speed of light, $c$, Planck's constant $h$, Newton's constant $G$, Boltzmann's constant $k$,  etc \cite{Duff:2001ba,Duff:2002vp}.  Any theory of gravitation and elementary particles can be characterized by a set of dimensionless parameters such as coupling constants
$\alpha_{i}$ (of which the fine-structure constant, $\alpha=e^2/\hbar c$, is an example),
mixing angles $\theta_{i}$ and mass ratios $\mu_{i}$.  To be concrete,
we may take $m_{i}{}^{2}=\mu_{i}{}^{2}\hbar c/G$ where $m_{i}$ is the
mass of the i'th particle.    For example, the standard model of particle physics coupled to gravity with a cosmological constant has 20 such dimensionless parameters as shown in Table I.  (There are more if massive neutrinos are included). See \cite{Rees} for other dimensionless numbers important for cosmology.

This difference is brought into focus most clearly when we entertain the possibility that the fundamental ``constants'' might be changing over cosmic time or from place to place in the present universe. (In practice this means that fundamental parameters are replaced by scalar fields $\phi^i$ in the Lagrangian whose equations of motion would typically admit space and/or time dependent solutions $\phi^i({\vec x}, t)$, as discussed in Section \ref{time}.)

For example, statements such as ``you cannot fix $c$ in theories where $c$ changes with time'' miss the point that $c$ changing in time is a matter of what units we choose, not what theory. To measure the speed of light we need a clock and ruler: if the distance between the notches on our ruler is the distance light travels between ticks of our clock then $c=1$ whatever our theory and will remain so until the cows come home. Astronomers who measure time in years and distance in light-years are doing exactly this. By contrast, as discussed in section \ref{views}, if the distance between notches on our ruler is the Bohr radius $ L_{B}=\hbar^{2}/m_{e}e^2$, say, and the time between ticks of our clock is the Bohr period $T_{B}=\hbar^{3}/m_{e}e^4$ then $c$ will have the same time variation (if any) as $1/\alpha$.  

Similarly, asking about the time variation of Newton's constant, or ${\dot G}/G$, is equally problematic when we can choose  $G=1$  units if we like. On the other hand, asking for example whether the number of protons needed  to meet the Chandrasekar bound has changed over cosmic history, is a sensible unit-independent question. 

These examples illustrate the general rule that all observers will agree on whether dimensionless numbers are changing in time but may disagree on dimensional ones \cite{Duff:2002vp}.  I claim no originality here. All this was well known to Dirac \cite{Dirac}, Jordan \cite{Jordan} and Dicke \cite{Dicke}, for example. It is curious therefore that this seemingly innocuous point of view should in fact be a source of much controversy, with intellectual heavyweights on both sides exchanging blows. We present some of these opposing views in section \ref{views} and respond accordingly \footnote{ 
Interestingly enough, one independent source with which I am in almost entire agreement is Wikipedia \cite{Phys.Const.}}.

%\newpage  
%\section{Tables and Figures}
 \begin{table}[ht]
% \begin{center}
\begin{tabular}{llllll}
Type&Number&\\
&&&&&\\
Yukawa coefficients for quarks ($u,d,c,s,t,b$) and  leptons ($e,\mu,\tau$)&9\\
Higgs $\mu_H=\sqrt{Gm_{H}{}^{2}/\hbar c}$ and coupling&2\\
 Three angles and a phase of the CKM matrix&4\\
 Phase for the QCD vacuum&1\\
 Coupling constants for the gauge group $SU (3) \times SU (2) \times U (1)$&3\\
 Cosmological parameter $G\hbar{\Lambda}/c^3$&1\\
 Massive neutrinos (model dependent)&7 or 8?\\
\end{tabular}
\label{1}
\caption{Dimensionless parameters of the Standard Model coupled to gravity with a cosmological constant $\Lambda$}
%\end{center}  
\end{table}

\section{Units, constants and conversion factors}
\label{constants}
\indent

\subsection{The Three Constants Party} 
\indent

As a young student of physics in school, I was taught that there
were three basic quantities in Nature: Length, Mass and
Time~\cite{Feather}.  All other quantities, such as electric charge or
temperature, occupied a lesser status since they could all be
re-expressed in terms of these basic three.  As a result, there were
three basic units: centimetres, grams and seconds, reflected in the
three-letter name ``CGS'' system (or perhaps metres, kilograms and
seconds in the alternative, but still three-letter, ``MKS'' system).

\begin{figure}\label{fig:p1} 
\begin{center}
\includegraphics[scale=0.5]{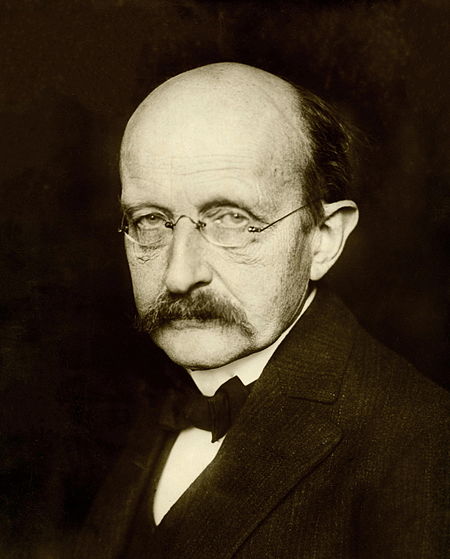}
\caption{\footnotesize{Max Planck} }
\label{Planck}
\end{center}
\end{figure}
Later, as an undergraduate student, I learned quantum mechanics,
special relativity and Newtonian gravity. In quantum mechanics, there
was a minimum quantum of action given by Planck's constant ${\hbar}$;
in special relativity there was a maximum velocity given by the
velocity of light $c$; in classical gravity the strength of the force
between two objects was determined by Newton's constant of gravitation
$G$. In terms of length, mass, and time their dimensions are
\begin{eqnarray}
[c]&=&LT^{-1}
\nonumber\\{}
[\hbar]&=&L^{2}MT^{-1}
\nonumber\\{}
[G]&=&L^{3}M^{-1}T^{-2}\,.
\end{eqnarray}
Once again, the number three seemed important and other dimensional
constants, such as the charge of the electron $e$ or Boltzmann's
constant $k$, were somehow accorded a less fundamental role.  This
fitted in perfectly with my secondary school prejudices and it seemed
entirely natural, therefore, to be told that these three dimensional
constants determined three basic units, first identified a century ago
by Max Planck \cite{Planck}, namely the Planck length $L_{P}$, the Planck mass
$M_{P}$ and the Planck time $T_{P}$:
\begin{eqnarray}
L_P&=&\sqrt{G\hbar/c^3}=1.616\times10^{-35}\,{\rm m}
\nonumber\\
M_P&=&\sqrt{\hbar c/G}=2.177\times10^{-8}\,{\rm kg}
\nonumber\\
T_P&=&\sqrt{G\hbar/c^5}=5.390\times10^{-44}\,{\rm s}
\label{Planck}
\end{eqnarray}

Yet later, researching into quantum gravity which attempts to combine
quantum mechanics, relativity and gravitation into a coherent unified
framework, I learned about the Bronshtein-Zelmanov-Okun (BZO)
cube~\cite{Bronshtein,Zelmanov,Okun:1996hz}, with axes $\hbar$, $c^{-1}$ and $G$, which neatly
summarizes how classical mechanics in the absence of gravity, non-relativistic quantum
mechanics, Newtonian gravity and relativistic quantum field theory can
be regarded respectively as the $({\hbar},c^{-1},G) \rightarrow 0$,
$(c^{-1},G) \rightarrow 0$, $({\hbar},c^{-1}) \rightarrow 0$, and $(G)
\rightarrow 0$ limits of the full quantum gravity.  Note, once again
that we are dealing with a three-dimensional cube rather than a square
or some figure of a different dimension.

Adherents of this conventional view of the fundamental constants of
Nature have been dubbed the ``Three Constants Party'' by Gabriele
Veneziano~\cite{Veneziano:1986zf}.  Lev Okun is their leader \cite{Duff:2001ba}.  For many years I was myself, I must confess, a card-carrying
member.   

\subsection{The Two Constants Party}
\indent

My faith in the dogma was shaken, however, by papers by Gabriele Veneziano
\cite{Veneziano:1986zf,Veneziano2,Veneziano3,Veneziano4}, self-styled leader of the
rebel Two Constants Party.  As a string theorist, Veneziano begins with
the two-dimensional Nambu-Goto action of a string.  He notes that,
apart from the velocity of light still needed to convert the time
coordinate $t$ to a length coordinate $x^{0}=ct$, the action divided
by $\hbar$ requires only one dimensional parameter, the string length
$\lambda_{s}^'$.
\begin{equation}
\lambda_{s}{}^{2}=\frac{\hbar}{cT_{}}\,,
\end{equation}
where $T_{}=1/2\pi c\alpha'$ is the tension of the string and
$\alpha'$ is the Regge slope. This is because the Nambu-Goto action
takes the form
\begin{equation}
\frac{S_{NG}}{\hbar}=\frac{1}{\lambda_{s}{}^{2}}{\rm Area}
\label{string}
\end{equation}
So if this were to describe the theory of everything (TOE), then the
TOE would require only two fundamental dimensional constants $c$ and
$\lambda_{s}$.
This claim led to many heated discussions in the CERN cafeteria
between Lev Okun, Gabriele Veneziano and myself. Weinberg \cite{Weinberg}
defines constants to be fundamental if we cannot calculate their
values in terms of more fundamental constants, not just because the
calculation is too hard, but because we do not know of anything more
fundamental.  This definition is fine, but it did not resolve the
dispute between Okun, Veneziano,  and me and we went round and round in circles.
\subsection{The Four Constants Party}
\indent
As a matter of fact, Planck \cite{Planck} himself treated temperature $\Theta$ on a par with length, mass and time; Boltzmann's constant $k$ 
\begin{equation}
[k]=ML^2T^{-2}\Theta^{-1}
\end{equation}
on a par with $\hbar$, $c$ and $G$ and  the Planck temperature:
\begin{equation}
\Theta_P=\sqrt{\hbar c^5/Gk^2}=1.416 \times 10^{32} K
\end{equation}
on a par with $L_P$, $M_P$ and $T_P$.

This case for including $k$ is argued forcefully in the book by Constantino Tsallis \cite{Tsallis}, so I would nominate him for leadership of the Four Constants Party.

\subsection{The Seven Constants Party}
\indent

%\newpage
\begin{table}[ht]
% \begin{center}
\begin{tabular}{llllll}
Date&Quantity&SI~Unit&Symbol&Dimension&Constant(2010)\\
&&&&&\\
1836&Mass&kilogram&kg&$M$&$\hbar=L^{2}MT^{-1}$\\
1836&Length&meter&m&$L$&$c=LT^{-1}$\\
1836&Time&second&s&$T$&$\Delta \nu (^{133} Cs)_{hfs}=T^{-1}$\\
1900&Current&ampere&A&$I$&$e=IT$\\
1954&Temperature&kelvin&$K$&$\Theta$&$k=ML^2T^{-2}\Theta^{-1}$\\
1954&Luminosity&candela&cd&$J$&$K_{cd}=J$\\
1971&Substance&mole&mol&$S$&$N_A=S^{-1}$\\
\end{tabular}
\label{2}
\caption{Seven ``basic'' SI units and their associated ``fundamental'' constants}
%\end{center}  
\end{table}
%\newpage
 \begin{table}[ht]
% \begin{center}
\begin{tabular}{llrll}
&&&\\
Constant&Symbol&Value&\\
&&\\
The speed of light in vacuum&$c$&$299~ 792~ 458$
%&Metre~per~second
&m~s$^{-1}$\\
The Planck constant &$h$&$6.626 ~06 \times 10^{-34}$
%& Joule~ second
&kg~m$^2$~s$^{-1}$\\
The elementary charge & $e$& $1.602~ 17 \times10^{-19}$
%& Coulomb
&A~s\\
The Boltzmann constant &$k$ &$1.380~ 6 \times10^{-23}$
%& Joule per kelvin
&kg~m$^2$s$^{-2}$K$^{-1}$\\
Caesium $133$ hyperfine splitting &$\Delta \nu (^{133} Cs)_{hfs}$&$ 9~ 192 ~631 ~770$
%& Hertz
&s$^{-1}$\\
The Avogadro constant & $N_A$ & $6.022 ~14\times 10^{23}$&mol$^{-1}$ \\
%& Reciprocal mol
The luminosity of  $540 \times 10^{12} $ hertz radiation&$K_{cd}$&$683$&cd\\
%&Lumen per watt
%   \hline
\end{tabular}
\label{3}
\caption{Seven ``basic'' dimensional SI constants.}
%\end{ruledtabular}
% \end{center}
\end{table}

\begin{figure}\label{fig:p1} 
\begin{center}
\includegraphics[scale=0.5]{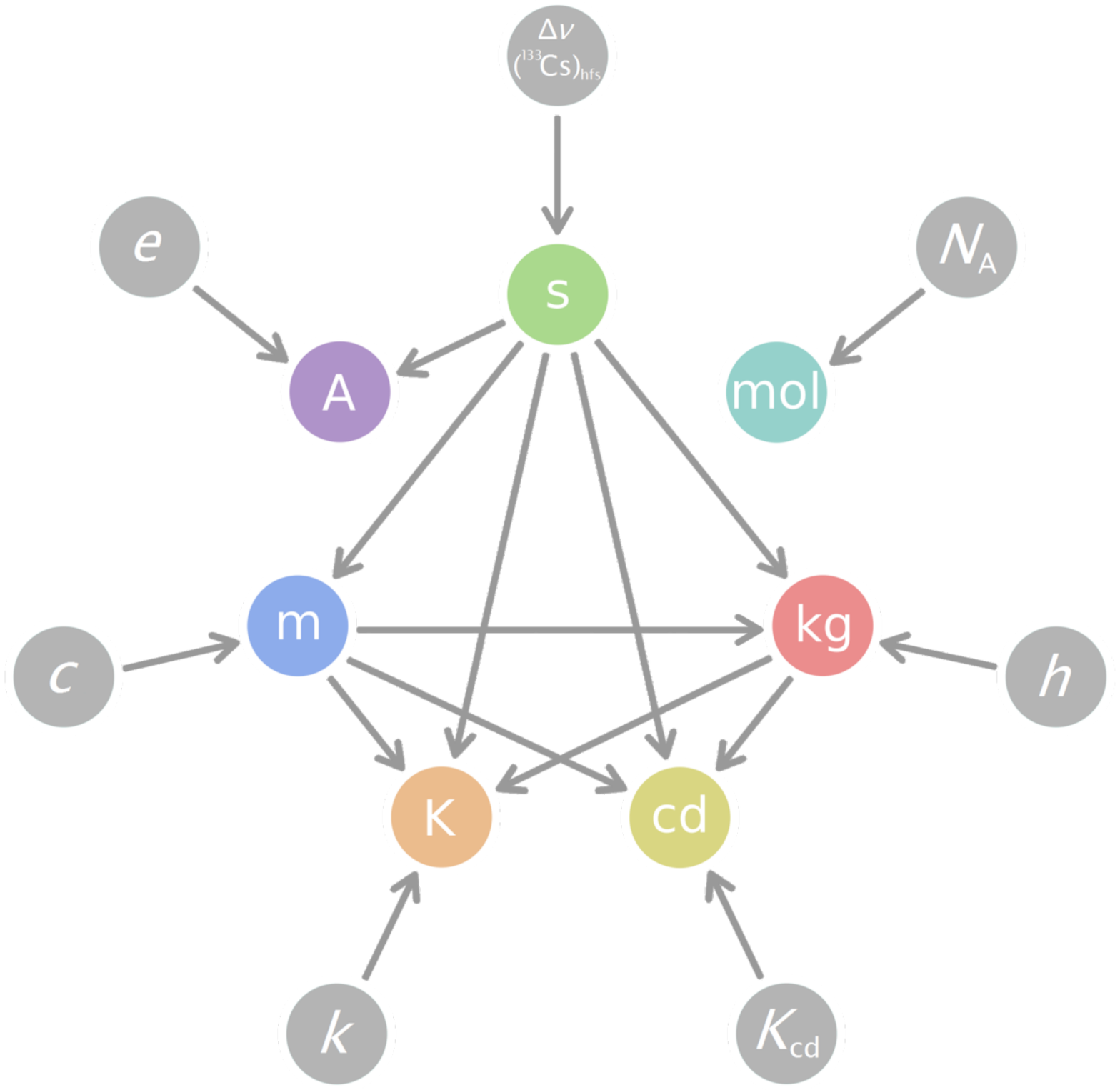}
\caption{\footnotesize{Dimensional constants in SI units} }
\label{Constants}
\end{center}
\end{figure}

Back in Texas, I continued these arguments at lunchtime
conversations with Chris Pope, Hong Lu and others.  There we eventually reached a consensus and joined what
Veneziano would call the Zero Constants Party.  Our attitude was basically that $\hbar$, $c$ and $G$ are nothing but
conversion factors e.g.\ mass to length, in the formula for the
Schwarzschild radius $R_{S}$
\[
R_{S}=\frac{2Gm}{c^{2}}\,,
\]
or  frequency to energy  
\[
E=\hbar \omega
\]
or mass to energy 
\[
E=mc^{2}
\]
no different from Boltzmann's constant, say, which relates 
temperature to energy 
\[
E=kT\,.
\]
As such, you may have as many so-called ``fundamental'' dimensional constants
as you like; the more different units you employ, the more
different constants you need. Indeed, no
less an authority than the \emph{Conf\'{e}rence G\'{e}n\'{e}rale
des Poids et Mesures}, the international body that administers the
SI system of units, adheres to what might be called the Seven
Constants Party, decreeing that seven units are ``basic'', as in Table 2,
 while the rest are ``derived''~\cite{NIST,SI}.  Accordingly, there are seven associated dimensional constants\footnote{One might argue that 7 units demand more than 7 conversion factors but I am reproducing  the CGPM presentation \cite{SI} as summarised by their diagram in Figure \ref{Constants}.} as in Figure 6.  To emphasise their role as mere conversion factors, in 2010 it was proposed to adopt the convention that they should all be fixed as in Table 3, rather than measured. They could have fixed $G$ also but decided not to. 
 
 Note, by the way, that even if these proposals for fixing $e$, $\hbar$ and $c$ are adopted this does not fix $\alpha$ because in SI units
 $\alpha=e^2/4\pi \epsilon_0 \hbar c$ where $\epsilon_0$ is the permittivity of free space. Varying $\epsilon_0$ cosmologies, anyone?

\subsection{The Zero Constants Party}
\indent

The attitude of the Zero Constants
Party is that the most economical choice is to use natural units
where there are no conversion factors at all.  Consequently, none
of these units or conversion factors is fundamental.  Since Okun, Veneziano and I were still unable to agree, we published a ``trialogue'' setting out our differences \cite{Duff:2001ba}.

\subsection{Not so fast}
\indent

The reason why we have so many different units in the first place is that, historically, physicists used
different kinds of measuring apparatus: rods, scales, clocks,
thermometers, electroscopes etc.  Another way to ask what is the
minimum number of basic units, therefore, is to ask what is, in
principle, the minimum number of basic pieces of apparatus.  Probably
Okun, Veneziano and  I would agree that $E=kT$ means that we can dispense with
thermometers, that temperature is not a basic unit and that
Boltzmann's constant is not fundamental.  Let us agree with Okun that
we can whittle things down to length, mass and time or rods, scales
and clocks.  Can we go further?  Another way to argue that the
conversion factor $c$ should not be treated as fundamental, for
example, is to point out that once the finiteness of $c$ has been
accepted, we do not need both clocks and rulers.  Clocks alone are
sufficient since distances can be measured by the time it takes light
to travel that distance, $x=ct$.  We are, in effect, doing just that
when we measure interstellar distances in light-years.  Conversely, we
may do away with clocks in favor of rulers.  It is thus superfluous to
have both length and time as basic units.

Moreover, we can do away
with rulers as basic apparatus and length as a basic unit by trading
distances with masses using the formula for the Compton wavelength
$R_{C}=h/mc$.  Indeed, particle theorists typically express length,
mass and time units as inverse mass, mass and inverse mass,
respectively.  Finally, we can do away with scales by expressing
particle masses as dimensionless numbers, namely the ratio of a
particle mass to that of a black hole whose Compton wavelength equals
its Schwarzschild radius.  So in this sense, the black hole acts as
our rod, scale, clock, thermometer etc.\ all at the same time.  In
practice, the net result is as though we set $\hbar=c=G=\cdots=1$ but
we need not use that language.

Wait a minute, I hear you say. The choice of length, mass and time as
the three basic units is due to Gauss~\cite{NIST}, so we could declare
him to be the founder of the Three Units Party, but not the Three Constants Party because this was
long before the significance of $c$ and $\hbar$ was appreciated. So does the number of basic units required 
and/or the number of fundamental dimensional constants depend on what point in history we pose the question?
For example, before relativity, length and time were treated differently and there was no constant $c$ to talk about.

This same point has been made eloquently by Wilczek \cite{Wilczek:2007iu} :

``In general, the more facts we allow ourselves to assume a priori, the fewer units, and the fewer fundamental constants, we need to introduce. But as a matter of principle we can only remove the Òfundamental constantÓ  by adopting a theoretical assumption. In general, by being bold we'll be economical, and appropriately ambitious, but we might be wrong.''

		I agree with Wilczek that the number of units is indeed theory dependent and changes with our understanding.  However, I differ by insisting that the number of {\it fundamental} dimensional constants was, is, and always will be zero.  Without relativity there was no constant $c$;  with relativity it is just a conversion factor. In measuring the distances along the $x$, $y$ and $z$ axes  in metres and distances along the $t$ axis in seconds one is reminded that sailors measure distances along the $x$ and $y$ axes 
in nautical miles and distances along the $z$ axis in fathoms. The number of metres to the second is no more a fundamental constant of nature than the number of fathoms to the nautical mile.  I am nevertheless grateful to an anonymous referee for the following observation:

{\it Originally, a nautical mile was defined as the distance one had to sail to move one minute of arc on the Earth's surface (presumably determined by astronomical observations), whereas a fathom was the distance an Old Frisian could spread his arms - convenient for measuring the length of wet rope used in estimating sea depth. Thus, in order to know the number of fathoms to the nautical mile one had to ask how many Old Frisians standing fingertip to fingertip are required to cover one minute of arc on the Earth's surface, which I assume, like the number of protons required to meet the Chandrasekar limit, is a good question. }

Moreover, Ellis and Uzan \cite{Ellis:2003pw}, with whom in other respects I find myself in agreement,  belong to Okun's Three Constants Party:

%{\it Ellis and Uzan \cite{Ellis:2003pw}:}
{\it How many such fundamental units are needed is still debated. To build on this debate (see Ref. \cite{Duff:2001ba} for different views), let us recall a property of the fundamental units of physics that seems central to us: each of these constants has acted as a Òconcept synthesizerÓ \cite{leblond,Lehoucq} , i.e. it unified concepts that were previously disconnected into a new concept. This for instance happens in the case of the Planck constant and the relation $E = \hbar \omega$, that can be interpreted not as a link between two classical concepts (energy and pulsation, or in fact matter and wave) but rather as creating a new concept with broader scope, of which energy and pulsation are just two facets. The speed of light also played such a synthesizing role by leading to the concept of space-time, as well as (with Newton's constant) creating the link, through the Einstein equations, between spacetime geometry and matter (see Refs. \cite{leblond,Lehoucq} for further discussion). These considerations, as well as facts on the number of independent units needed in physics \cite{Uzan:2002vq,Duff:2001ba} tend to show that three such quantities are needed. It also leads, when investigated backward, to the concept of the cube of physical theories \cite{Okun:1996hz}.}

I agree that, historically speaking, the speed of light played a very different role than the speed of sound, for instance.  My objection is that synthesizing the concepts of space $x,y,z$  and time $t$ is equivalent to the statement that the laws of physics are invariant not merely under the rotation group $O(3)$ but under the Lorentz group $O(3,1)$.  But landlubbers synthesised the sailor's $x,y$ and the sailor's $z$ by noticing that the laws of physics are invariant not merely under $O(2)$ but $O(3)$. So by the same reckoning the number of fathoms to the nautical mile is just as much a ``concept synthesiser'' as the speed of light. Yet as the anonymous referee continues:

{\it Only later was it noticed that the laws of physics could be made invariant under O(3) by redefining both the nautical mile and the fathom. Thus, I feel that Wilczek's observation might be given more weight.}

I give the final words of this section to J-M.~Levy-LeBlond~\cite{leblond}: 

\indent
{\it This, then, is
the ordinary fate of universal constants: to see their nature as
concept synthesizers be progressively incorporated into the implicit
common background of physical ideas, then to play a role of mere unit
conversion factors and often to be finally forgotten altogether by a
suitable redefinition of physical units.}
%%%%%%%%%%%%%%%%%%%%%%%%%%%%%%%%%%%
 
\section{\bf Time variation of fundamental ``constants''}
\label{time}
\indent

\subsection{An example}
\label{example}
The claim \cite{Webb} that the fine-structure
constant, $\alpha$-the measure of the strength of the electromagnetic
interaction between photons and electrons-is slowly increasing over cosmological
time scales has refuelled an old debate about varying fundamental
constants of nature. In our opinion \cite{Duff:2002vp}, however, this debate
has once again been marred by a failure to distinguish between  dimensionless
and dimensional constants.  An example of this
confusion is provided in  \cite{Davies:zz}, where it is claimed that
 ``As
$\alpha=e^{2}/{\hbar}c$, this would call into question which of these
fundamental quantities are truly constant''.  

By consideration of
black hole thermodynamics, the authors conclude that theories with
decreasing $c$ are different from (and may be favored over) those with 
increasing $e$.  Here we argue that this claim is operationally 
meaningless, in the sense that no experiment could tell the 
difference, and we replace it by a meaningful one involving just 
dimensionless parameters.

The authors of \cite{Davies:zz} point out that the entropy $S$ of a
non-rotating black hole with charge $Q$ and mass $M$ is given by
\begin{equation}
S=\frac{k\pi G}{\hbar c}[M+\sqrt{M^{2}-Q^{2}/G}]^{2}
\label{entropy}
\end{equation}
They note that decreasing $c$ increases $S$ but increasing $e$, and hence
$Q$, decreases $S$.  It is then claimed, erroneously in our view, that
black holes can discriminate between two contending theories of
varying $\alpha$, one with varying $c$ and the other with varying $e$.

Let us define the dimensionless parameters $s$, $\mu$ and $q$ by
$S=sk\pi$, $M^{2}=\mu^{2}\hbar c/G$ and $Q^{2}=q^{2}\hbar c$.  The 
mass ratio $\mu$ 
will depend on the fundamental dimensionless parameters of the theory 
$\alpha_{i}$, $\theta_{i}$ and $\mu_{i}$, but the details need not 
concern us here. We shall shortly give a thought-experimental definition of $s$, $\mu$
and $q$ that avoids all mention of the unit-dependent quantities $G$,
$c$, $\hbar$, and $e$. 
Shorn of all its irrelevant unit dependence, therefore,
the entropy is given by
\begin{equation}
s= [\mu+\sqrt{\mu^{2}-q^{2}}]^{2}
\label{entropy1}
\end{equation}
If we use the fact that the charge is quantized in units of $e$, namely
$Q=ne$ with $n$ an integer, then $q^{2}=n^{2}\alpha$, but we prefer not
to mix up macroscopic and microscopic quantities in (\ref{entropy}).

The unit dependence of the claim in \cite{Davies:zz} that black holes
can discriminate between varying $c$ and varying $e$ is now evident. 
For example, consider the first three units in Table 4.
\noindent
In Planck units \cite{Planck,Duff:2001ba}
\begin{equation}
\hbar=c=G=1~~~~~~~\alpha=e^2~~~~~~~M^2=\mu^{2}
\label{Planck1}
\end{equation}
In Stoney units \cite{Stoney,Duff:2001ba}
\begin{equation}
c=e=G=1~~~~~~~\alpha=1/{\hbar}~~~~~~~M^{2}•=\mu^{2}•/\alpha
\label{Stoney1}
\end{equation}
\begin{figure}\label{fig:p1} 
\begin{center}
\includegraphics[scale=0.7]{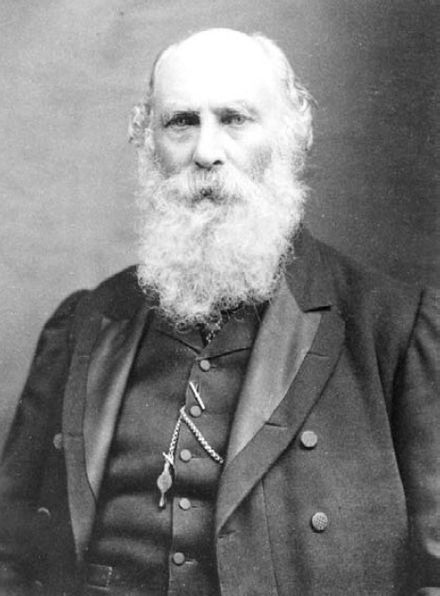}
\caption{\footnotesize{George Stoney} }
\label{Stoney}
\end{center}
\end{figure}
In Schr\"odinger units \cite{Duff:2002vp}
\begin{equation}
\hbar=e=G=1~~~~~~~\alpha=1/c~~~~~~~M^{2}•=\mu^{2}•/\alpha
\label{Schrodinger1}
\end{equation}
\begin{figure}\label{fig:p1} 
\begin{center}
\includegraphics[scale=1.5]{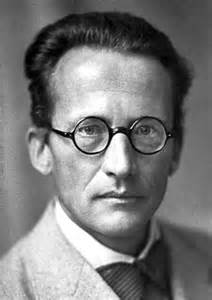}
\caption{\footnotesize{Erwin Schr\"odinger} }
\label{Schrodinger}
\end{center}
\end{figure}

In all three units (and indeed in {\it any} units), the dimensionless 
entropy ratio $s$ is the same as given by (\ref{entropy}).  To 
reiterate: assigning a change in $\alpha$ to a change in $e$ (Planck) 
or a change in $\hbar$ (Stoney) or a change in $c$ (Schr\"odinger) is 
entirely a matter of units, not physics.  Just as no experiment can 
determine that MKS units are superior to CGS units, or that degrees 
Fahrenheit are superior to degrees Centigrade, so no experiment can 
determine that changing $c$ is superior to changing $e$, contrary to 
the main claim of Davies et al \cite{Davies:zz}.

So far we have discussed units in which $c$, ${\hbar}$, and $e$ may vary.  In Dirac units, shown in Table 4, $G$ may vary.
\begin{equation}
c=e=m_{e}=1~~~~~\hbar=1/\alpha~~~~~G=\mu_{e}{}^{2}/\alpha
~~~~M^{2}=\mu^{2}/\mu_{e}{}^{2}
\label{Dirac1}
\end{equation}

\begin{figure}\label{fig:p1} 
\begin{center}
\includegraphics[scale=1.0]{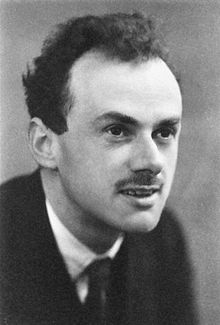}
\caption{\footnotesize{Paul Dirac} }
\label{Dirac}
\end{center}
\end{figure}

Once again, the entropy is the same as given by (\ref{entropy}).
So there is no such thing as a varying $G$ {\it theory}, only 
varying $G$ {\it units}. This is familiar from string theory \cite{GSW}
where the 
string tension $T$ is related to $G$ via dilaton and moduli fields 
which may possibly vary in space and time.  In Einstein units, $G$ is
fixed 
while $T$ may vary, whereas in string units $T$ is fixed while $G$ may 
vary.

For the sake of completeness, we also define Bohr length, mass, 
time and charge as in Table 4, which have an obvious atomic definition as the Bohr
radius etc.  Note that these units are independent 
of $G$ and $c $.  They are obtained from Schr\"odinger units by the
replacement 
$G \rightarrow G \alpha/\mu_{e}{}^{2}$. In Bohr units
\begin{equation}
\hbar=e=m_{e}=1~~~~~c=1/\alpha~~~~~G=\mu_{e}{}^{2}/\alpha~~~~
~~~~M^{2}=\mu^{2}/\mu_{e}{}^{2}
\end{equation}
\begin{figure}\label{fig:p1} 
\begin{center}
\includegraphics[scale=1.5]{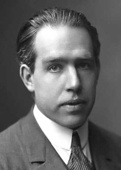}
\caption{\footnotesize{Niels Bohr} }
\label{Bohr}
\end{center}
\end{figure}

Can we give a thought-experimental (as
opposed to purely mathematical) meaning to these length, mass,
time and charge units? Interestingly, the extreme charged  black hole solution provides the answer.  Its Schwarzschild radius is
\begin{equation}
R_{S}=\frac{GM}{c^{2}}+\sqrt{\frac{G^{2}M^{2}}{c^{4}}
-\frac{GQ^{2}}{c^{4}}}
\end{equation}
and its Compton wavelength is
\begin{equation}
R_{C}= \frac{h}{Mc}.
\end{equation}
In the extreme case, moreover, we have
\begin{equation}
R_{S}=\frac{GM}{c^{2}}~~~~~~Q^{2}=GM^{2}
\end{equation}
$L_{P}$, $M_{P}$, $T_{P}$ and $Q_{P}$ may now be thought-experimentally
defined 
without reference to any fundamental constants as
the Schwarzschild radius, mass, characteristic time and charge of a black hole whose Schwarzschild radius equals its Compton wavelength divided
by $2\pi$.  Thus $s$, $\mu$ and $q$ count the number of times $S$, $M$
and 
$Q$ exceed the entropy, mass and charge of such an black hole.  
Similarly, Stoney length, mass, time and charge are the corresponding 
quantities for an extreme charged black hole whose charge is the charge on the electron.  
A thought-experimental definition of Schr\"odinger $L^{2}$ is
the Bohr 
$L^{2}$ scaled down by Dirac's large number (the ratio of 
electromagnetic to gravitational forces $e^{2}/Gm_{e}^{2}$) with 
similar definitions for $M^{2}$ and $T^{2}$.
%\newpage
 \begin{table}[ht]
 \begin{center}
%\begin{ruledtabular}
\begin{tabular}{llllllll}
%\hline
%\hline
&&&&&\\
&&Planck&Stoney&Schr\"odinger&Dirac&Bohr\\
&&&&&\\
$L_{}{}^{2}$&&$G\hbar /c^3$&$Ge^{2}/c^4$&$G\hbar^{4}/e^6$&$e^{4}/m_{e}{}^{2}c^4$&$\hbar^{4}/m_{e}{}^{2}e^4$&\\
 $M_{}^{2}$&&$\hbar c/G$&$e^{2}/G$&$e^{2}/G$&$m_{e}{}^{2}$&$m_{e}{}^{2}$&\\
$T_{}{}^{2}$&&$G\hbar/c^5$&$Ge^{2}/c^6$&$G\hbar^{6}/e^{10}$&$e^{4}/m_{e}{}^{2}c^6$&$ \hbar^{6}/m_{e}{}^{2}e^8$&\\
 $Q_{}{}^{2}$&&$\hbar c$&$e^{2}$&$e^{2}$&$e^{2}$&$ e^{2}$&\\
 &&&&\\
 %Property%&``natural''&$\hbar \rightarrow \alpha \hbar$&$c \rightarrow \alpha c$&$G \rightarrow G\alpha m_{e}{}^{2}$\\
% &no~$e$&no~$\hbar$&no~$c$&no~$G$,no~$\hbar$&no~$G$ \\
%$\alpha$&&$e^2$&$1/\hbar$&$1/c$&$1/\hbar $&$1/c$\\
%$\mu^2$&&$M^2$&$M^2/\hbar$&$M^2/c$&$GM^2/\hbar$&$GM^2/c$\\
%$q^2$&&$Q^2$&$Q^2/\hbar$&$Q^2/c$&$Q^2/\hbar$&$Q^2/c$
\end{tabular}
\label{4}
\caption[]{Length, Mass, Time and Charge in Planck, Stoney, Schr\"odinger, Dirac and Bohr units. Planck units are independent of $e$, Stoney units are independent of $\hbar$ and so we call the units independent of $c$ `` Schr\"odinger'', even  though he may never have used them.}
%\end{ruledtabular}
 \end{center}
\end{table}

%\newpage

\subsection{How can ``constants'' vary?}

How in practice would our equations accommodate varying ``constants"? Merely replacing a constant C by C(t)
would not pass muster as a viable theory. The only sensible way is to introduce scalar fields $\phi^i$ into our Lagrangian. The  fundamental constants would then appear as (dimensionless combinations of ) vacuum expectation values of these scalars whose equations of motion would typically admit time dependent solutions $<\phi^i(t)>$. Such fields appear naturally in higher-dimensional Kaluza-Klein theories where components of the metric tensor in the extra dimensions correspond to scalars in four-dimensional space-time. One might even suppose that the ultimate theory starts out with no parameters at all in its Lagrangian and that all the familiar parameters of particle physics and cosmology, possible  plus some new ones, would emerge as vacuum expectation values of scalar fields. Indeed M-theory \cite{DuffM} fulfills this requirement. Alternatively, one might simply postulate a tensor/scalar theory directly in four dimensions as Brans and Dicke \cite{Brans} did. Replacing parameters by scalar fields as the only sensible way to implement time
varying constants of Nature is also emphasized in \cite{Duff:2002vp,Veneziano:1986zf,Veneziano2,Veneziano3,Veneziano4,
Ellis:2003pw,Damour:2002vu,Barrow:2002mj}.

Of course this raises the problem that in comparing theory to experiment it is the theory with varying scalar fields we must use and its predictions for cosmology, black holes and particle physics will not be of the standard type. For example, although M-theory admits charged black hole solutions of the kind discussed in Section \ref{example}, and although it also admits solutions with varying $\alpha$ and $\mu$, I know of no solutions that do both simultaneously. So as discussed in  \cite{Duff:2002vp},  the question of whether black holes could discriminate between contending theories with different variations of $\mu$ and $q$, as opposed to $c$ and $e$ as claimed by Davies et al \cite{Davies:zz}, is obscured by these difficulties. This problem is not unique to string theory; it is shared by all such tensor/scalars theories.  This point was also made by Barrow in a subsequent paper entitled ``Unusual Features of Varying Speed of Light Cosmologies'' \cite{Barrow:2002mj}. Puzzlingly, however, instead of addressing my objections to the whole concept of varying speed of light, Barrow criticises  \cite{Duff:2002vp}  for failing to have solved this scalar problem.  Unlike Barrow \cite{Barrow:2002mj}, I do not believe the uncertainties surrounding scalar fields exonerate those papers claiming to have measured the time variation of dimensional constants.  For example, according to the Planck collaboration \cite{Ade:2014lua}:

{\it Any variation of the fundamental physical constants, and more particularly of the fine structure constant, $\alpha$, or of the mass of the electron, $m_e$, would affect the recombination history of the Universe and cause an imprint on the cosmic microwave background angular power spectra. 

Another aspect of the variation of fundamental constants, which has been much discussed in the literature (see e.g., Dicke \cite{Dicke}; Duff \cite{Duff:2002vp}; Uzan \cite{Uzan:2010pm}; Narimani et al. \cite{Narimani:2011rb}), is that only dimensionless combinations of constants can really be measured. Because of this, many previous studies have focussed on the parameter $m_e/m_p$. We have checked that for the physics of recombination our consideration of $m_e$  is entirely equivalent to variation of $m_e/m_p$.   Hence our study of constraints on ($\alpha,m_e$) is consistent with arguments that the only meaningful variations are dimensionless ones.} 

I am not quite sure what ``We have checked that for the physics of recombination'' means here.  It seems they have simply adopted a system of units in which the proton mass is fixed i.e. Dirac units with $m_e$ replaced by $m_p$. They go on to say:

{\it However, the situation would be more complicated if we were to consider additional constants, and in particular $G$.  As already stated, $G $ enters the Friedmann equation, and so even if one considers a dimensionless ratio, such as $Gm_p/\hbar c$, there are still complications over whether the cosmological framework is even self-consistent, in addition to whether the cosmological perturbations might evolve differently. This can only be done within the context of specified theories of modified gravity.}

A variable $m_e/m_p$ also requires a deviation from the standard model of particle physics so similar theoretical caveats apply. In either case it is not clear to me what are they claiming to have measured.
If the theoretical uncertainties prevent them from knowing what is being measured, so be it.

Two of the authors, Moss and Scott, together with Narimani \cite{Narimani:2011rb} have elsewhere broken ranks and emphasised he need for a ``dimensionless cosmology'' :

{\it We have not explored all examples in the literature of constraints on $G$, and we have not exhaustively assessed each and every paper. However, it is clear that at least some of the published discussions involving $G$ rather than $Gm_p/\hbar c$ are in fact constraints on a different combination of parameters than asserted by the authors.
We believe that cosmologists should get their gravitational house in order, and speak only of $Gm_p/\hbar c$. }

\subsection{Summary}
In summary, it is operationally meaningless \cite{Duff:2001ba} and
confusing to talk about time variation of arbitrary unit-dependent
constants whose only role is to act as conversion factors.  For
example, aside from saying that $c$ is finite, the statement that
$c=3\times 10^{8}~m/s$, has no more content than saying how we convert
from one human construct (the meter) to another (the second).  Asking
whether $c$ has varied over cosmic history (a question unfortunately
appearing on the front page of the {\it New York Times} \cite{Glanz}, 
in {\it Physics
World} \cite{PW1}\footnote{To its credit, {\it Physics World} also published
the dissenting view \cite{PW2}.}, in {\it New Scientist} \cite{NS1,NS2,NS3}, in {\it Nature} \cite{Davies:zz}
and on {\it CNN} \cite{CNN}) is like asking whether the number of litres
to the
gallon has varied.

\section{Opposing views}
\label{views}
In this section, we present a selection of opposing views to be found in the literature. Quotes are in italics, followed by our responses.

\subsection{Moffat \cite{Moffat}}
\indent

{\it In a recent article, Duff \cite{Duff:2002vp} has asserted that dimensional constants such as $c$, $\hbar$ and $G$ ``are merely human constructs whose number and values differ from one choice of units to the next and which have no intrinsic physical significanceÓ .  Of course, as long as these dimensional constants remain constants, then we can set them equal to unity, and treat them as a means to change units. However, once we postulate that these constants are no longer really ÒconstantÓ but vary in space and time, we can no longer assert that they are just Òhuman constructsÓ that allow us to change from one set of units to another.}

{Response}: As we have seen, even if dimensionless constants are
changing in time, nothing stops us from using Planck units with
$c=\hbar=1$ and time varying $e$, Stoney units with $c=e=1$ and
time varying $\hbar$ or Schr\"odinger units with $\hbar=e=1$ and time
varying $c$.

{\it  It seems clear that whether you vary $e$, $\hbar$ or $c$ will have very different consequences for physics. Such consequences can be detected and measured and from these results, we can decide which ÒdimensionalÓ constant of the three involved is varying, even though the effects of a varying $\alpha$ appear to be falsely ÒhiddenÓ in the variation of either $e$, $\hbar$ or $c$. Considering the variation of $\alpha$ in isolation from the rest of physics and not taking into account the variation of either $e$, $h$ or $c$ individually seems an unacceptable approach to the problem. It is conceivable that varying the charge $e$ could lead to a theory that somehow could be re-written as a theory in which $e$ is kept fixed and $c$ is varied, but this would lead to a strange and very complicated revision of all of physics.}

{ Response}: On the contrary, it is nothing more than switching from Planck units to Stoney units.

{\it Dirac was one of the first physicists to suggest that, in connection with his theory of large numbers, fundamental dimensional constants may vary in time during the expansion of the universe. Indeed, he considered that Newton's gravitational constant $G$ varied with time. If one so wishes, one can consider the measurable quantity, ${\dot G}/G$, in which the only dimensional quantity that enters the formula is the time $t$ and $t$ is measured by standard clocks.
}

{ Response}:  In his 
seminal paper \cite{Dirac} Dirac says: ``The fundamental constants of 
physics, such as $c$ the velocity of light, $h$ the Planck constant, 
$e$ the charge and $m_{e}$ the mass of the electron, and so on, 
provide for us a set of absolute units for measurement of distance, 
time, mass, etc.  There are, however, more of these constants than are 
necessary for this purpose, with the result that certain dimensionless 
numbers can be constructed from them.'' The phrase ``more of these 
constants than are necessary'' is crucial.  Those who insist on 
counting the dimensional constants in a theory as well as the 
dimensionless ones will always have more unknowns than equations.  
This redundancy is nothing but the freedom to change units without 
changing the physics.  In Einstein-Maxwell-Dirac theory, for example, 
one could imagine units in which (at least) five dimensional 
constants, are changing in time: $G$, $e$, $m_{e}$, $c$, 
$\hbar$\ldots, but only two dimensionless combinations are necessary: 
$\mu_{e}{}^{2}=Gm_{e}{}^{2}/\hbar c$ and $\alpha=e^{2}/\hbar c$.

Dirac then notes that the {\it dimensionless} ratio of electromagnetic and
gravitational forces $e^{2}/Gm_{e}^{2}$ is roughly the same order of
magnitude as the {\it dimensionless} ratio of the present age of the
universe 
$t$ and the atomic unit of time $e^{2}/m_{e}c^{3}$.  He makes it clear
that 
equating these two numbers leads to a time-varying $G \sim t^{-1}$ only
in the 
``atomic units'', denoted ``Dirac'' in Table 4.

{\it  If we assume that $h$ varies in time with $c$ kept constant, this would produce detectable effects in atomic spectra but it would not obviously alter quantum mechanics at a fundamental level, nor would it require a revision of special relativity.}

{ Response:} I agree but you are simply describing a change in $\alpha$ in Stoney units. The same statement could be made in a unit-independent way.

\subsection{Davies \cite{Davies}}
\indent

{\it Where 
we differ substantially from Duff, and where it seems clear he is 
wrong, is in his claim that theories in which dimensional constants 
vary with time are  ``operationally meaningless.'' Such theories have 
existed in the literature, and specific observational tests been 
suggested and carried out, at least since Dirac's theory of varying $G$.}

{ Response}: I agree that Davies et al are the latest in a long line of
authors making such claims, but Dirac was not one of them.   

{\it Some
theories of fundamental physics, e.g. the Hoyle-Narlikar theory of
gravitation, were explicitly designed to incorporate an additional gauge
freedom (in that case, conformal invariance) to enable one to transform at
will between different systems of units, without changing the physics,
whilst including cosmological time variations of constants. }

{ Response}: The freedom to choose MKS units, say, over CGS units 
requires no symmetry of the fundamental theory but is merely one of
human 
convention.  The same is true of choosing changing $c$ units
 over changing $e$ units. 
  
  {\it The speed of light is more than an electrodynamic
parameter: it
describes the causal structure of spacetime, and as such is relevant to
all of
physics (for example, the weak and strong interactions), not just
electrodynamics. }

{Response}: What is relevant for the strong, weak and 
electromagnetic interactions is the special theory of relativity, i.e
invariance under the Poincare group of spacetime transformations. 
The mathematics of the Poincare group ($x'{}^{\mu}=
\Lambda^{\mu}{}_{\nu}x^{\nu}+ a^{\mu}$)
can get along just fine without $c$.

 Let us suppose that we have a generally covariant and 
locally Lorentz invariant theory of gravity with scalar fields, and that
time varying $\alpha$ is implemented by a time-dependent scalar field 
solution. 
This background will not exhibit global Lorentz invariance, but this is
no 
different than a time-dependent Friedman-Robertson-Walker cosmology 
which is not Lorentz-invariant either. Alternatively, we might 
imagine a phase transition from one Lorentz-invariant vacuum to another 
in which the dimensionless constants, such as $\alpha$, change abruptly.
Whatever 
the symmetries, they 
will be the same whether we use varying $c$ units or some other units. 
Moreover, none of this conflicts with Einstein's general 
covariance, contrary to certain claims in the literature and in the
media.

{\it A variation of $c$ cannot be mimicked in all such respects
by a
change in $e$.  More obviously, one can imagine measuring the speed of
light in the
laboratory tomorrow and obtaining a different value from today. That is
clearly operationally meaningful.}

{Response}: This common fallacy can be eliminated by thinking carefully
about how one would attempt to measure $c$ in a world in which
dimensionless constants such as $\alpha$ and $\mu$ are changing in
time.  First take a ruler with notches one Planck length apart and a
clock with ticks one Planck time apart.  Next measure the speed of
light in vacuum\footnote{If the experiment is performed in a medium, 
or a time-dependent gravitational field, one would have to factor out
the 
effects of the refractive index, or $\sqrt{g_{xx}/g_{tt}}$.  After all, 
light slows down when passing through a piece of glass, but no-one is 
suggesting that this produces an increase in $\alpha$.} by counting 
how many notches light travels in between ticks.  You will find the 
answer $c=1$.  You may repeat the experiment ad infinitum  
and you will always find $c=1$!  Repeat the experiment using Stoney 
length and Stoney time, and again you will find $c=1$.  But if the 
notches on your ruler are one Schr\"odinger length apart and the ticks 
on your clock one Schr\"odinger time apart, you will find $c=1/\alpha$ 
and $c$ will now have the same time dependence as $1/\alpha$.  Once 
again we see that the time dependence of $c$ is entirely 
unit-dependent.  Similar remarks apply to the measurement of any other 
dimensional quantity.
Measuring the speed of light with a ruler whose notches are one 
Bohr length apart and a clock whose ticks are one Bohr time apart will 
again result in $c=1/\alpha$.  As discussed in \cite{Albrecht}, Bohr 
units are used when measuring $c$ using an atomic clock, which is most 
sensitive to a variation of $\alpha$.  A pendulum clock, on the other 
hand, is more sensitive to the variation of $\mu_{i}$.  So when you 
think you are measuring a dimensional quantity, you are really 
measuring dimensionless ones.

{\it So this is an issue of semantics and mathematical
elegance, not science.}

{Response}: The failure to tell the difference between changing
units and changing physics is more than just semantics. It brings to mind the 
old lady who, when asked by the TV interviewer whether she believed in  
global warming, responded: ``If you ask me, it's all this changing 
from Fahrenheit to Centigrade that's causing it''.

\subsection{Magueijo \cite{Magueijo}} 
\indent

{\it We start by discussing the
physical meaning of a varying $c$, dispelling the myth that the
constancy of $c$ is a matter of logical consistency...
In discussing the physical meaning of a varying speed of light,
I'm afraid that Eddington's religious fervor is still with
us \cite{Ellis,Duff:2002vp} \footnote{It is curious that  {\it Nature} rejected \cite{Duff:2002vp} on the grounds that Davies, Moffat and Magueijo were right,  but published  \cite{Ellis} on the grounds that they were wrong. } . ``To vary the speed of light is
self-contradictory'' has now been transmuted into ``asking whether
$c$ has varied over cosmic history is like asking whether the
number of liters to the gallon has varied'' \cite{Duff:2002vp}. The implication is that the constancy of
the speed of light is a logical necessity, a definition that could
not have been otherwise. This has to be naive. For centuries the
constancy of the speed of light played no role in physics, and
presumably physics did not start being logically consistent in
1905. Furthermore, the postulate of the constancy of $c$ in
special relativity was prompted by experiments (including those
leading to Maxwell's theory) rather than issues of consistency.
History alone suggests that the constancy (or otherwise) of the
speed of light has to be more than a self-evident necessity.}

{\ Response: } In fact my remark implies no such logical necessity. 
It merely means
that the variation or not of dimensional numbers like $c$ (as opposed to
dimensionless numbers like the fine-structure constant) is a matter
human convention, just as the variation or not in the number of liters
to a gallon is a matter of human convention. In neither case is it
something to be determined by experiment but rather by one's choice of
units.Ê So there is no such thing as a varying $c$ `theory' only 
varying $c$ `units'.Ê For example, in units where time is measured in years and
distance in light-years, $c=1$ for ever and ever, whatever your theory!

As a matter of fact, the number of liters per gallon varies as 
one crosses the Atlantic. Similarly, as mentioned in the Introduction, in 1983 the Conference Generale des Poids 
et Mesures changed the number of meters per second, i.e the value of c.  Relativity 
survived intact!

		{\it If $\alpha$  is seen to vary one cannot say that all the dimensional parameters that make it up are constant. Something Ð $e$, $h$ $c$, or a combination thereof Ð has to be varying. The choice amounts to fixing a system of units, but that choice has to be made.  A possible way to evade this argument is to say that physical theories should only refer to directly measurable dimensionless parameters \cite{Duff:2002vp}, a view I label fundamentalism. }

{ Response:} This is the crucial point. The main thrust of the present paper is to argue that the laws of physics require no such choice. Thus
  
 Bad question: Is $m_e$ or $m_p$ varying in time?
 
 Good question: Is $m_e/m_p$ varying in time?
 
 Bad question: Is $c$ or $h$ or $e$ varying in time?
 
 Good question: Is $\alpha$ varying in time?
 
 Bad question: Is $G$ or $m_p$ or $h$ or $c$ varying in time?
 
 Good question: Is the number of protons required to meet the Chandrasekar limit varying in time?
 
 Good question: Is the number of orbits required before the perihelion of Mercury returns to it original position varying in time?

``Good '' means all observers will agree on the answer; ``bad'' means the answer is unit-dependent and will generically depend on the rods, clocks and scales used to make the measurement.

\subsection{Barrow \cite{Barrow:2002,Barrow2002}}
\indent

{\it  These results also raise the question: which of $e$, $\hbar$ and $c$ might be responsible for any observed change in $\alpha$ and what operational meaning should be attributed to such a determination? Undoubtedly, in the sense of \cite{Duff:2001ba}, one has to make an operationally ÒmeaninglessÓ choice of which dimensional constant is to become a dynamical variable. }

{ Response:} So far, so good

{\it  Yet, in practice this choice is never arbitrary; it is clearly dictated by simplicity once the detailed dynamics of the theory have been established. Here, we argue that the dynamics have unambiguous observational implications: a combination of experiment and simplicity therefore selects one member of a dimensionless combination $\alpha$ of dimensional constants $e$, $\hbar$ and $c$  to which we should preferentially ascribe its space-time variation. We will present a number of clear experimental tests which can distinguish rival theories of $\alpha$ variation which are expressed through explicit change in $e$ or $c$. }
  
{Response:} In my opinion, the rival theories in question are distinguishable, not because you have preferentially ascribed the variation of alpha to different dimensional constants, but because they are different theories with different lagrangians. This is obscured by choosing to call one a varying $c$ theory and the other a varying $e$ theory. 

{\it An important lesson we learn from the way that pure numbers like $\alpha$ define the World is what it really means for worlds to be different. The pure number we call the fine structure constant and denote by $\alpha$ is a combination of the electron charge, $e$, the speed of light, $c$, and Planck's constant, $h$. At first we might be tempted to think that a world in which the speed of light was slower would be a different world. But this would be a mistake. If $c$, $h$, and $e$ were all changed so that the values they have in metric (or any other) units were different when we looked them up in our tables of physical constants, but the value of $\alpha$ remained the same, this new world would be observationally indistinguishable from our World. The only thing that counts in the definition of worlds are the values of the dimensionless constants of Nature. If all masses were doubled in value you cannot tell, because all the pure numbers defined by the ratios of any pair of masses are unchanged .}

{ Response:}  Of course I agree. Yet you have written about a dozen papers with ``varying $G$'' or ``varying $c$'' in their title.  For example \cite{Barrow:2002mj,Barrow:1999is,Clifton:2005xr}. Isn't this confusing?

\subsection{Gibbons \cite{Gibbons:2014zla}}

\indent

{\it Until the development of Quartz, Ammonia and Caesium clocks,
time measurements were astronomical, and the default assumption was
that with respect to those units,  Newton's law of gravity was     
independent of time.  The most economical assumption was 
then that the rate of atomic processes are governed by the same
units \cite{Kelvin}. Thus the times which enter  Kepler's law and
Schr\"odinger's equation are the same and coincide with
those that enter Maxwell's equations. In which case  the  three `` fundamental 
constants of physics'' $G,\hbar$ and $c$ would indeed be constants
and (Planck) units could be adopted in which they  
be taken without loss of generality to equal unity \cite{Planck}.
However the constancy of all three constants has been questioned,  most
notably $G$ by, among others, Dirac \cite{Dirac}, Jordan \cite{Jordan} , 
Brans and Dicke \cite{Brans}.  }

{ Response:} On the contrary, they were questioning the constancy of dimensionless numbers, and well aware that
the variation of dimensional ones devolves upon the choice of units.

{\it One may also question the constancy of $\hbar$ and $c$  
but the evidence against any time variability appears to be so strong
that in this paper I shall assume that they are indeed constant.}

{ Response:}  Again this is a matter of human conventions, not evidence.

{\it In principle, the other various constants of the standard model,
could vary with time, but current limits appear to be extremely stringent
and so in this paper I shall assume that such things
as  the ``fine structure constant'' are indeed constant.
If this is not true, we would for example, have to introduce
Stoney time \cite{Stoney,Barrow1983}. }

{Response:}  Again I disagree: the whole virtue of the dimensionless parameters of the standard model, such as the fine structure constant, is that they are the same in anybody's units.

{\it A linguistic purist
might justifiably object to  this oxymoron and even point out that
it only makes meaningful physical sense to say that dimensionless ratios of 
physical quantities may vary with time. However 
since the construction of the requisite dimensionless quantities 
is little more than an elementary undergraduate exercise I shall 
not trouble the reader by spelling it out in detail.  }

{\it Response: } As the above responses make clear, our differences extend beyond mere linguistic purity.

\subsection{Reasenberg \cite{Reasenberg}}
\indent

{\it Dirac has investigated the cosmological consequences of the large number hypothesis. It was noted even earlier that it is possible to combine physical constants to create dimensionless numbers that generally differ from unity by at most a few orders of magnitude. One of these is the ratio of the electric to the gravitational force between an electron and a proton; it is close to $10^{40}$. Another is the age of the universe in units of atomic time; it is close to $10^{40}$. Finally there is the mass of the visible universe expressed in proton masses; it is close to $10^{(40 \times 2)}$. The hypothesis is that this coincidence is a message, not an accident; perhaps these quantities are related by some 
time-invariant constants... }

{Response: }  So far, so good.

{\it If that is correct, then one or more of the the ``constants'' used to make each of the large numbers
must be time-varying. The original description was that the gravitational ``constant'' was the most like candidate for the time variable.}

{\it Response:} This is putting things backwards, in my opinion. Changing a question about the dimensionless numbers into one about dimensional numbers such as $G$, is changing a question about the laws of nature into one about human conventions.

\subsection{Uzan \cite{Uzan:2010pm}}
\indent

{\it The numerical values are given in the Planck system of units defined by the requirement that the numerical value of $G$, $c$ and $ \hbar$ is 1 in this system of units, (Page 7).}

{ Response: } Good, so ${\dot G}/G$=0?

{\it  Monitoring the orbits of the various bodies of the Solar system offers a possibility to constrain deviations from general relativity, and in particular the time variation of $G$....An early analysis of this data assuming a Brans-Dicke theory of gravitation gave that $|{\dot G}/G| \leq 3 \times 10^{-11} yr^{-1}$,(Page 71).}

{Response:} A good example of how confusion can arise by asking unit-dependent questions. For the most part, however, Uzan's views and mine are in agreement.

\subsection{Copi, Davis and Krauss \cite{Krauss}}

{\it Big Bang Nucleosynthesis (BBN) can provide, via constraints on the expansion rate at that time, limits on $G$. We find that
\[
 -3 \times 10^{-13} yr^{-1}\leq({\dot G}/G)_{today} \leq 3 \times 10^{-13} yr^{-1}
 \]
 }
{Response:}  Here we take the opportunity to make the point that dimensionless 
ratios such as $\Delta G/G$, $\Delta e/e$ and $\Delta c/c$ are every 
bit as unit-dependent as their dimensional counterparts $\Delta G$, 
$\Delta e$ and $\Delta c$. An obvious example is again provided by 
units in which time is measured in years and distance in light-years.
Here $c=1$ and $\Delta c/c$=0, whatever your theory. Similar remarks 
apply to $\Delta G/G$. As discussed in Section \ref{example}, it is guaranteed to 
vanish in Planck units (\ref{Planck1}), for example, but might vary in Dirac 
units (\ref{Dirac1}). By contrast, $\Delta \alpha/\alpha$ is 
unit-independent.

 It is the variation of the parameters in Table I
that may be constrained by the astrophysical data presented in 
\cite{Krauss}, not $\Delta G$ nor even $\Delta G/G$.
\subsection{Quinn \cite{Quinn}}
\indent

 {\it For a scientist, and a former director of the International Bureau of Weights and Measures (BIPM) in Paris such as myself, the imprecision in $G$ is irritating. Moreover, there is a solid scientific case for sorting it out. }
 
{ Response: }  So far, so good.
 
{\it The search for a theory of quantum gravity that is consistent with quantum electrodynamics is perhaps the most active field of theoretical physics.  One day, we may have to test such theories by comparing the values of G that they predict with the real thing Ñ so we need an accurate experimental value.}

{ Response: } No theory will ever predict a human convention such as G.
What might logically be predicted by theory one day are the dimensionless constants of nature, such as those in Table 1. Being pure numbers, they are the same in anybody's units. Robust experiments are indeed required to pin them down more accurately, though so far theory has no explanation for any of them.  
 
It could be that some, or maybe even all, of these numbers are simply accidents of our place in the landscape of universes known as the multiverse. If this were the case, they will never be predicted by theory either, though we can narrow down their possible values by noting that if they were only slightly different we wouldn't be here to study them.

\subsection{Contemporary Physics \cite{CP}}
\indent

{\it Contemporary Physics, Instructions to Authors :} Authors must adhere to SI units .

{Response:} I did my best.

\section{Conclusions}
\label{conclusions}\label{secM7}

The number, current values, and possible time variation of the dimensionless constants appearing
in the laws of physics is a legitimate subject of physical enquiry. They are worthy of the title ``fundamental''.
By contrast, the number, current values and possible time variation of the dimensional constants, such as
$h$, $c$, $G$,\ldots\ are quite arbitrary human constructs.  There is nothing magic about
the choice of number: two or three or seven or...Their numerical values are subjective, differing from one choice of units to the next. Accordingly it is matter of convention whether they are something we measure or something we define to be fixed.  Consequently, none of these dimensional constants is fundamental. 

\section{Acknowledgements}
  I would like to thank my ``adversaries'' in Section \ref{views} (several of whom are personal friends) for their thought-provoking contributions to this debate. If they wish to present a case for the defence of views I have disagreed with in this paper, I would be delighted to hear them.  I am also grateful to Lev Okun and Gabriele Veneziano for first arousing my interest in this topic and to  Chris Pope, Hong Lu and Constantino Tsallis for stimulating conversations. This work is supported by the STFC under rolling grant ST/G000743/1.
  
  \bigskip

  \bigskip
ADDED NOTE: I am grateful to my colleague Stanley Deser for making  the following three points: (1) The modern Wilsonian effective field theory approach to particle physics and the renormalisation group place greater emphasis on the running of the coupling constants with energy than on their value at some particular energy; (2)  $G$ is ``dynamical'' and should not be lumped together with $h$ and $c$ which are ``kinematical''; (3)  It is mixing apples and oranges to talk about time-varying constants of the standard model since they are defined to be t-independent.

Response: (1) Yes I agree this is an important point (and one made independently by Frank Close).  For the fine structure constant, for example, $\alpha(Q^2=0)\sim 1/137$ but  $\alpha(Q^2=M_W{}^2)\sim 1/128$. This well-established energy dependence should not be confused with the more speculative cosmic time dependence of $\alpha(Q^2=0)$ discussed in the text; (2) OK, but does not affect my conclusions; (3) Agreed and as I argue in section 3.2,  the standard model would have to be modified by replacing parameters by scalar (or pseudoscalar) fields e.g. by replacing the theta angle by an axion.

\section{The author} 
Michael Duff received his PhD in theoretical physics from Imperial College London in 1972.  He returned to Imperial College as a member of staff in 1979, taking leave of absence to visit the Theory Division at CERN from 1984 to 1987 when he became Senior Physicist. He took up his professorship at Texas A\&M University in 1988 and was appointed Distinguished Professor in 1992. In 1999 he became Oskar Klein Professor of Physics at the University of Michigan and was elected to serve as first Director of the Michigan Center for Theoretical Physics 2000-2005. In 2005 he returned once more to Imperial where he has served as Principal of the Faculty of Physical Sciences and is currently Abdus Salam Professor of Theoretical Physics. He was elected Fellow of the Royal Society in 2009. Michael's interests lie in unified theories of the elementary particles, quantum gravity, supergravity, Kaluza-Klein theory, superstrings, supermembranes, M-theory and quantum information theory.\newpage
\begin{figure}\label{fig:p1} 
\begin{center}
\includegraphics[scale=1.0]{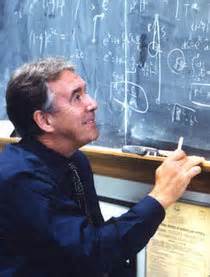}
\caption{\footnotesize{The Author} }
\label{Duff}
\end{center}
\end{figure}

\end{document}